# InGaN/GaN µLED SPICE modelling with size dependent ABC model integration

Anis Daami*[a], François Olivier[a]
[a]Université Grenoble Alpes, CEA-LETI, Minatec Campus, III-V Lab., Grenoble, France

## ABSTRACT

The need of high brightness micro-displays in portable applications dedicated to mixed and/or virtual reality has drawn an important research wave on InGaN/GaN based micro-sized light emitting diodes (µLEDs). We propose to use a SPICE modelling technique to describe and simulate the electro-optical behavior of the µLED. A sub-circuit portrayal of the whole device will be used to describe current-voltage behavior and the optical power performance of the device based on the ABC model. We suggest an innovative method to derive instantaneously the carrier concentration from the simulated electrical current in order to determine the µLED quantum efficiency. In a second step, a statistical approach is also added into the SPICE model in order to apprehend the spread on experimental data. This µLED SPICE modelling approach is very important to allow the design of robust pixel driving circuits.

## 1. INTRODUCTION

Many research efforts have been carried out and a huge amount of interesting results can be found in literature dealing with the comprehension of electro-optical behavior of InGaN/GaN micro-light emitting diodes[1,2,3] (µLEDs). Indeed, the growing interest to use these devices, especially in display and/or portable modules, is still at its very first years as no proven application can be found in the market today. An important reason is the high amount of µLEDs needed, that have to be efficiently similar to achieve high quality µ-displays[4,5,6].

The design of these latter, is based on the combination of a pixelated array and generally a CMOS active matrix to drive the µLEDs at their best quantum efficiency. Therefore, robust circuit design is a mandatory keystone to ensure a good functioning of these applications. Indeed process fabrication spread, even brought to a minimum, is inherent to any technology. As known and well employed in robust technologies, such as in microelectronics platforms, CMOS designs, are dependent on SPICE[7] device models.

Some work has been done on devices like organic photodiodes (OPDs) or organic light emitting diodes (OLEDs) to capture through a SPICE model their electro-optical behaviors[8,9,10]. Some efforts towards modelling LEDs behavior can be found in literature. Nevertheless, to our knowledge, these papers generally present an analytical modelling study[11,12,13] of the device or a TCAD based model[14]. Therefore it has become urgent and almost mandatory to develop such models for devices like LEDs to simulate, understand and enhance the opto-electrical properties of µ-displays.

Based on our latest experimental published data on blue emitting µLEDs[15,16], we report in this paper a complete SPICE model construction. This model aims to a complete description of the electro-optical behavior of µLEDs through a large size range. We demonstrate the use of a simple mathematical resolution technique in the simulation core to calculate the injected carrier density from the ABC model equations. Consequently, the LED external quantum efficiency (EQE) calculus becomes straight forward and is emulated inside the model itself, allowing the accurate determination of the emitted optical power. Furthermore, we show the importance of the statistical approach in this type of modelling to allow the design of robust pixel driving circuits dedicated to µ-display applications.

* anis.daami@cea.fr

## 2. LED OPTO-ELECTRICAL BEHAVIOR MODELLING

### 2.1 The sub-circuit approach

We show on figure 1 the equivalent electrical circuit used to model the electrical behavior of the LED. It consists of a SPICE LEVEL 1 ideal diode (D), a series resistor (RS) and a parallel resistor (RP). This sub-circuit has two input pins representing the LED anode and cathode. The light output behavior is described by a virtual current-controlled current source (F). This latter is biased between two nodes VN and VP that are transparent to the final user. The current value ($I_F$) flowing through this virtual source F emulates the optical output power ($P_{opt}$) of the LED and is defined as:

$$I_F = P_{opt} = \frac{I_{LED}}{q} \cdot EQE \cdot \frac{hc}{\lambda} \qquad (1)$$

where $I_{LED}$, $\lambda$ and EQE are respectively, the current flowing through input pins (anode to cathode), the emitted wavelength (460 nm in our case) and the external quantum efficiency of the LED. The reader will easily recognize the universal constants h, c and q. We will focus on the determination of EQE in section 2.3.

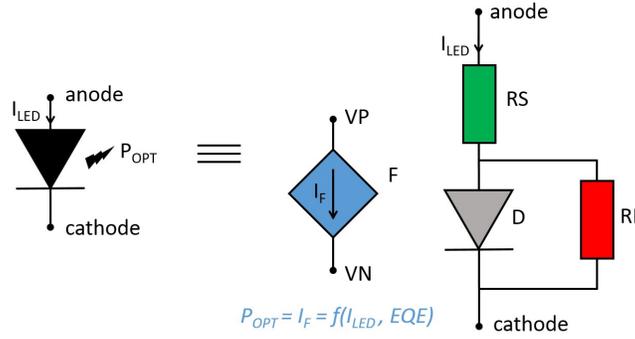

Figure 1. Equivalent electrical circuit implemented to model the LED electro-optical behavior. The components RS, RP and D are used to describe the current-voltage characteristic of the LED. The F current source is a virtual user-transparent component to mimic the optical output power of the LED.

### 2.2 Electrical parameters adjustments and size effect

Figure 2 shows current-voltage measurements carried out on vertical square-shaped LEDs of different size, ranging from 10µm to 500µm. In order to adjust our SPICE model to these electrical characteristics, we have chosen to put to nil the internal series resistance of the SPICE LEVEL1 ideal diode D. The value of the implemented external resistor RS has been extracted for each LED size. The whole set permitted to derive an empirical area-dependent relationship equation as follows:

$$RS = \frac{S0}{S} \cdot \left(RS_0 + RS_V \cdot e^{-\alpha(V-V0)}\right) \qquad (2)$$

where S0 and S define the reference LED area (500µm width), and the modelled one, respectively. Knowing that $S \leq S0$, it is evident that RS will present a greater value for small devices.

Furthermore, equation (2) can be interpreted as a sum of two series resistances contributions where the first term ($RS_0$) is fixed and represents the common part. The second term is voltage dependent and is seen as a non-linearity deviation of RS. Indeed, the series resistance is a representation of all the physical obstacles that injected carriers have to overcome to flow through the device. Generally, anode and cathode contacts show a deviation from ideal ohmic behavior when bias is pushed to high values. Process fabrication can also generate defects and recombination centers preventing a smooth carrier flow. For all these reasons the RS parameter value cannot be considered as a fixed value. Finally, parameter V0 is inferred as a threshold voltage from which RS non-linearity begins to act, whereas α is a smoothing adjustment parameter.

On the other hand, the value of the parallel resistance RP could not be adjusted for small geometries due to the limitation in our measurement instrumentation. This also demonstrates that our devices have extremely weak leakage currents at low voltage, proving a high process quality. To simplify the model and avoid any dysfunction during simulations, we have decided to give RP a fixed value, satisfying the lowest LED leakage current we have measured. Evidently, this choice underestimates the leakage of large geometries. Nevertheless, a light emitting device is often biased far from the leakage region above its threshold voltage, therefore this modelling hypothesis is acceptable.

Beside the RS and RP adjustments, we have tuned the saturation current (IS) and ideality coefficient (η) parameters in the SPICE LEVEL1 diode model (D) to fit the low leakage and intermediate bias regions, far from the high voltage zone. Both parameter values seemed to be geometry independent.

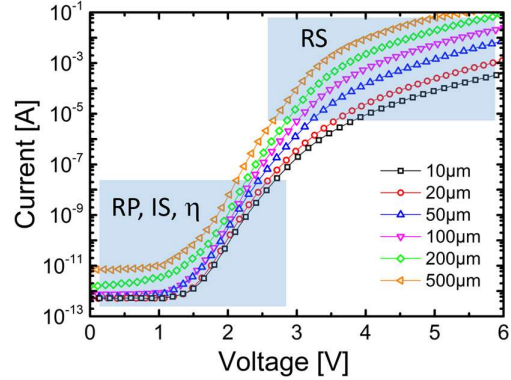

Figure 2. Current-voltage characteristics of different square LEDs. Width ranges from 10µm to 500µm. Shaded areas show where the SPICE model parameters, RS, RP, IS and η were adjusted. Note that the leakage current saturates due to a limitation in our measurement instrumentation.

## 2.3 Optical quantum efficiency and ABC model

The quantum efficiency of LEDs has thoroughly been described by the well documented ABC model in literature. We have recently suggested a slight modification to take into account size in the ABC model[9]. This modification principally affects parameter A related to the Shockley Read Hall (SRH) non-radiative recombination. Indeed, we have demonstrated that A is proportional to the ratio perimeter/surface (P/S) of the LED. Parameters B and C, respectively linked to radiative and Auger non-radiative recombinations, are shown to be geometry independent. They represent the epitaxy material quality. Taking into account the electron charge density q and the quantum well thickness t, the current density $J_{LED}$ and the internal quantum efficiency (IQE) of the LED are expressed as follows:

$$J_{LED} = q \cdot t \cdot \left(A \cdot \frac{P}{S} \cdot n + B \cdot n^2 + C \cdot n^3\right) \tag{3}$$

and

$$IQE = \frac{B \cdot n^2}{A \cdot \frac{P}{S} \cdot n + B \cdot n^2 + C \cdot n^3} \tag{4}$$

Here, n is the injected carrier density flowing through the device at a given voltage. Knowing the light extraction efficiency (estimated LEE ≅ 0.14 by ray tracing simulations), the external quantum efficiency EQE is then determined from IQE in this way:

$$EQE = IQE \cdot LEE \tag{5}$$

We propose in this SPICE modelling methodology to integrate parameters A, B and C as input entities of the device, at the same level as its geometrical inputs, namely its perimeter P and surface S. It is then obvious that if the injected carrier density n is known, one will not have any difficulty to estimate the quantum efficiency of the LED. Nevertheless, any SPICE simulation only allows an electrical current determination.

Our original submitted approach is then to derive instantaneously the value of n from the simulated electrical current. Indeed, equation (3) shows that for a given simulated current density $J_{LED}$, the calculus of n is brought to the roots determination of a 3rd degree polynomial ($a \cdot x^3 + b \cdot x^2 + c \cdot x + d = 0$). This resolution is not straightforward as one might think. Indeed, depending on the polynomial coefficients a, b, c and d different mathematical resolution methods have been proposed. We have decided to implement in the core of the LED model library, an easy and common technique called the Cardano's method[17, 18]. This method allows a rapid calculus of all real solutions if the 3rd degree polynomial coefficients are real, which is the case in the ABC model.

Therefore, for each applied voltage value, the current density is simulated using the adjusted electrical LED SPICE model, and the resolution of equation (3) to determine the carrier density n is simultaneously carried out. Subsequently, the quantum efficiency is determined using equations (4) and (5).

**2.4 A statistical approach**

To provide a better robustness of the LED model, we have chosen to introduce a statistical modelling approach to allow the use of Monte Carlo SPICE simulations. At first approximation, the choice has been made to principally affect the injected carrier density into the LED. Therefore, we implemented Gaussian distributions on each recombination parameter A, B and C. Average parameter values have been fixed to those used in the typical model (from our latest experimental results[9]) and a given standard deviation around 15% has been fixed for all 3 parameters. This random choice has no particular reason, except the aim to apprehend the quantum efficiency spread due to the parameters cited above. Thus, for each random Monte Carlo draw of A, B and C, the same calculus of n, IQE and EQE will be carried out as stated in section 2.3. In this special statistical configuration, it is clear that the simulated LED current will not change from a draw to another, as no variation has been introduced in the LED model parameters. Evidently, the further step would be, to evaluate the mutual effects of current variation, and recombination parameters distributions together on the LED quantum efficiency spread.

## 3. SIMULATION RESULTS

**3.1 Electrical characteristics**

We show on figure 3, the simulated current-voltage characteristics superimposed to measurement ones for the different LED sizes. It is clear that the implemented SPICE model describes well the electrical behavior, on a large voltage range. As discussed above, the current at low voltage ($0 \leq V \leq 1$) is a little bit underestimated for large geometries due to the choice of a constant value for RP accounting for LED leakage (not shown on figure 3). Despite this little discrepancy, the developed sub-circuit electrical SPICE model approach shows a high quality adjustment over the large variety of LED sizes.

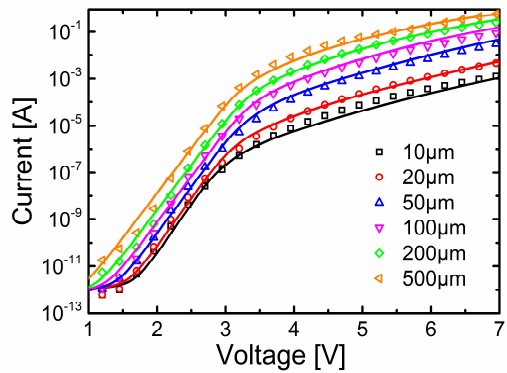

Figure 3. SPICE model simulated current-voltage characteristics (lines) versus experimental data (symbols) of different LED sizes. A good adjustment is observed on a large voltage range for all geometries.

## 3.2 Quantum efficiency

Figure 4 shows external quantum efficiency EQE versus current density characteristics for all studied LED geometries. For the sake of clarity, experimental curves (figure 4 (a)) and simulated ones (figure 4 (b)) are not presented on the same graph. The first observation to come out is that the experimental EQE aspect is well represented by simulation. Indeed, first we observe the optical threshold shifting towards higher current density levels for small geometries. Moreover, the maximum efficiency $EQE_{max}$ drops when size diminishes as perceived on experimental data. Additionally, the common sinking of EQE known as the ''droop'' effect is also present on simulated graphs. Yet, looking closely to figure 4, some differences exist between measurements and simulations. Indeed the EQE optical threshold appears at lower current densities for experimental data. Besides, the $EQE_{max}$ drop versus LED size is somewhat exaggerated in the simulated EQE graphs. These differences are mainly due to the ABC model hypothesis, assuming a current-density $J_{LED}$ relationship to carrier density n expressed as stated in equation (3), which in reality is an empirical model based approximation. Nevertheless, for SPICE simulations it is sufficient to have the essence of the electro-optical behavior of InGaN/GaN based LEDs.

Another point worth noting is the break appearing on simulated curves whatever the LED size. This is inherent to the resolution method we have used to determine the carrier density n. Indeed, Cardano's method for the roots determination of a 3$^{rd}$ degree polynomial, is highly dependent on the polynomial coefficients. This dependence implies 2 kinds of calculus subject to coefficient values. Knowing that one coefficient is directly related to current density $J_{LED}$, this break shows the transit between both styles of the polynomial root resolution.

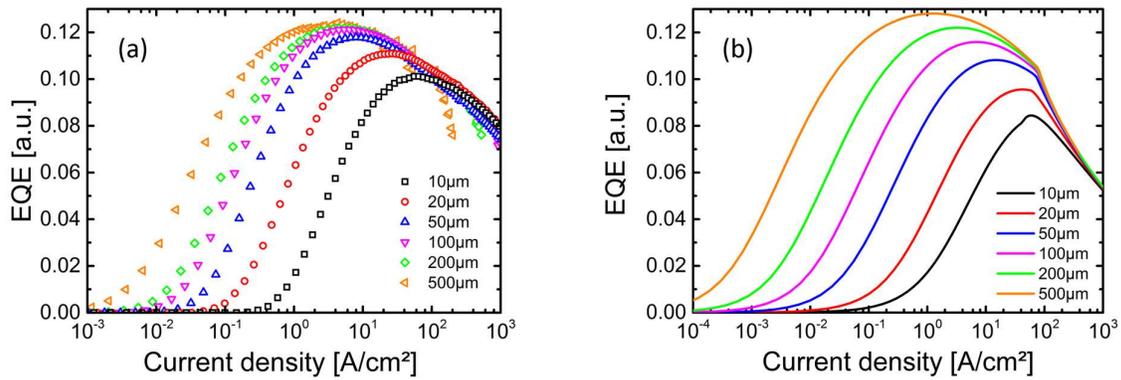

Figure 4. External quantum efficiency versus current density for different LED sizes: experimental data behavior shown in symbols (a) are well described by the simulated SPICE model (b). Some differences are yet observable between measurements and simulation, due to the ABC model approximation on the LED current modelling.

## 3.3 Spread and recombination parameters effect

To apprehend the effect of recombination mechanisms on the LED optical behavior, we have run Monte Carlo simulations for a 50µm width LED using the statistical approach we implemented in our SPICE model. We have focused our attention on the spread of $EQE_{max}$. Simulations results for 1000 runs are shown on figure 5. The first observation is that the maximum external quantum efficiency shows a Gaussian like distribution (figure 5 (a)) centered on an average value $\langle EQE_{max} \rangle = 0.11$. with a spread deviation at $3\sigma = 0.02$. To seek the effect of the random A, B and C value draws on quantum efficiency, all extracted $EQE_{max}$ values are presented on three different cloud graphs versus all drawn values of recombination parameters, respectively. Even though, no special correlation between recombination parameters has been used (random draws of A, B and C), we observe awaited evolutions of $EQE_{max}$ versus each related recombination parameter. Indeed, we observe a tendency of having a better quantum efficiency when a higher parameter B value is drawn (figure 5 (c)), whatever are A and C values. This trend is also observable when A (figure 5 (b)) or C (figure 5 (d)) show lower values independently of the two other parameters. At last, we think that this kind of cloud representations of Monte Carlo simulations can also help understand and quantify the spread in quantum efficiency of µLEDs dedicated for display applications. Indeed high resolution µLED arrays contain a huge number of single µLEDs that will have to be controlled efficiently by an active matrix which also has its own spread. Therefore an InGaN/GaN LED SPICE model is mandatory to evaluate and design robust µLED displays.

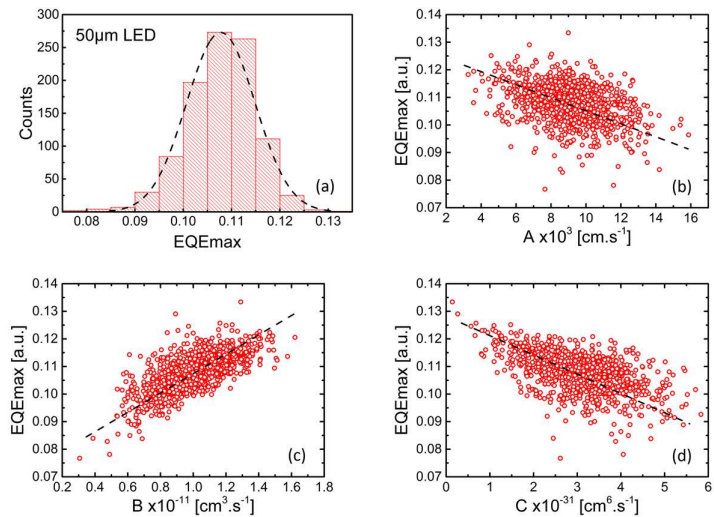

Figure 5. 1000 run Monte Carlo simulation results of a 50µm width LED. (a) Histogram Gaussian like distribution of $EQE_{max}$. Cloud graphs in (b), (c) and (d) represent $EQE_{max}$ versus SRH, radiative and Auger recombination related parameters A, B and C, respectively. Awaited trends of $EQE_{max}$ are observed versus each parameter showing the robustness of our implemented statistical approach in the SPICE model.

## 4. CONCLUSION

An InGaN/GaN LED SPICE model taking into account known size effects is demonstrated. A sub-circuit description of the LED permitted an accurate modelling of the electrical behavior. The use of the ABC model recombination related parameters with a simultaneous mathematical resolution during current simulations allowed a precise determination of the external quantum efficiency versus current density. Moreover, a first simple statistical approach based on Gaussian distributions applied to recombination parameters show that this kind of SPICE modelling can help understand and ameliorate pixel designs dedicated to high resolution displays that are aimed to applications for virtual, augmented or mixed reality.

## ACKNOWLEGEMENTS

The authors acknowledge partial fundings from the European Union's Horizon 2020 VOSTARS research and innovation programme under grant agreement No 731974 and H2020 HILICO European project (H2020- JTI-CS2-2016-CFP04-SYS-01-03, Grant No. 755497).